\documentclass[namedreferences,fleqn]{SolarPhysics}
\usepackage{graphicx}

\begin{document}
\bibliographystyle{aa}
\begin{article}
\begin{opening}

\title{A New Look at Mode Conversion in a Stratified Isothermal Atmosphere}

\author{A.M.D.~\surname{MCDOUGALL} and A.W.~\surname{HOOD}}

\runningauthor{A.~M.~D.~\surname{MCDOUGALL} and A.~W.~\surname{HOOD}}
\runningtitle{MODE CONVERSION IN A STRATIFIED ISOTHERMAL ATMOSPHERE}

\institute{School of Mathematics and Statistics, University of St Andrews, KY16 9SS, UK\email{dee@mcs.st-and.ac.uk}\\}

\date{Received 10 January 2007; accepted 5 March 2007}

\begin{abstract}

Recent numerical investigations of wave propagation near coronal magnetic null points \cite{McLaughlin2006} have indicated how a fast MHD wave partially converts into a slow MHD wave as the disturbance passes from a low-$\beta$ plasma to a high-$\beta$ plasma.  This is a complex process and a clear understanding of the conversion mechanism requires the detailed investigation of a simpler model.  An investigation of mode conversion in a stratified, isothermal atmosphere, with a uniform vertical magnetic field is carried out, both numerically and analytically.  Contrary to previous investigations of upward propagating waves \cite{Zhugzhda1982a,Cally2001}, this paper studies the downward propagation of waves from a low-$\beta$ to high-$\beta$ environment.  A simple expression for the amplitude of the transmitted wave is compared with the numerical solution.  

\end{abstract}

\end{opening}

\section{Introduction}
\label{Introduction} 

Mode conversion in the solar atmosphere has been a problem of interest for many years.  It is important when the plasma $\beta$, the ratio of the gas pressure to the magnetic pressure, is close to unity; this is the region where the sound and Alfv\'en speeds are of comparable magnitude.  Outside this area the modes are effectively decoupled, with one behaving like an acoustic wave and the other displaying a strong magnetic nature.  Understanding the processes that occur as waves propagate through this $\beta\approx1$ layer may in turn tell us the modes involved in coronal heating, and alert us to what we should be searching for observationally.  

Stein (1971) presented a full and detailed study of mode conversion across a density step, using the dispersion relations and boundary conditions to calculate transmission and reflection coefficients for fast, slow, and Alfv\'en waves.  The conditions under which the different modes are coupled to each other were also described.  Zhugzhda (1979) and Zhugzhda and Dzhalilov (1981; 1982a,b) tackled the problem analytically, focusing on an isothermal atmosphere, stratified by gravity, and permeated by a uniform vertical magnetic field.  A solution was found in terms of Meier-G functions, specifically $G_{24}^{12}$, allowing transformation coefficients to be found for waves propagating from high $\beta$, low in the atmosphere, to low $\beta$, high in the atmosphere.  This study was extended by Cally (2001), who found that this solution could be expressed fully in terms of the equivalent hypergeometric $_2F_3$ functions.  Reflection, transmission and conversion coefficients were again found for waves incident from below, with another set of coefficients discovered in addition to those found by Zhugzhda and Dzhalilov (1982a).  Our work uses the same simple one-dimensional set-up as these authors, but instead of studying the propagation from high $\beta$ to low $\beta$ we examine propagation in the opposite direction.  This study takes the form of a numerical investigation backed up by analytical work.  Similarly, McLaughlin and Hood (2006) examined the behaviour of a fast magnetoacoustic wave travelling from a low-$\beta$ to a high-$\beta$ plasma, in this case as it approaches a magnetic null point.  As the wave front crosses the $\beta\approx1$ region, the fast wave can be seen to partially convert into a slow wave.  

Further numerical simulations have been carried out in two dimensions \cite{Rosenthal2002,Bogdan2003}, two-and-a-half dimensions \cite{Khomenko2006}, and in three dimensions \cite{Carlsson2006}.  All of these simulations found that the magnetic canopy (defined to be the isosurface where $\beta=1$) is a crucial factor in mode conversion, along with the attack angle for the incident wave.  This is supported analytically by Cally (2006), and by Bloomfield {\it et al.~}(2006) who found observational evidence that the dominant wave mode changes from the fast mode to the slow mode as the waves propagate across the magnetic canopy.  However, due to the complex nature of the numerical simulations it is difficult to disentangle individual physical effects.  The simplicity of the model used in our investigation allows a much clearer picture to be built up, greatly aiding the understanding of the processes involved in mode conversion.  This in turn will aid the interpretation of results from more complex models and simulations.  

The paper has the following outline.  In Section~\ref{Basic Equations} we introduce the MHD equations in the form that we shall be using them, detailing the assumed $x$-dependence and the non-dimensionalisation.  The numerical simulations are discussed in Section~\ref{Num Sims}, particularly in relation to their agreement with previous analytic studies.  Finally we summarise our findings in Section~\ref{conc.sec}.  

\section{Basic Equations}
\label{Basic Equations}

Throughout we work with the standard ideal MHD equations, namely
\begin{eqnarray*}
\rho\left[\frac{\partial}{\partial t}+{\bf v}\cdot{\bf\nabla}\right]{\bf v}&=&-{\bf\nabla}p+\frac{1}{\mu}\left({\bf\nabla}\times{\bf B}\right)\times{\bf B}+\rho{\bf g},\\
\frac{\partial{\bf B}}{\partial t}&=&{\bf\nabla}\times\left({\bf v}\times{\bf B}\right),\\
\frac{\partial\rho}{\partial t}+{\bf\nabla}\cdot\left(\rho{\bf v}\right)&=&0,\\
\left(\frac{\partial}{\partial t}+{\bf v}\cdot{\bf\nabla}\right)p&=&\frac{\gamma p}{\rho}\left(\frac{\partial}{\partial t}+{\bf v}\cdot{\bf\nabla}\right)\rho,\\
p&=&\frac{R\rho T}{\widetilde{\mu}},
\end{eqnarray*}
where $\rho$ is the mass density, ${\bf v}$ is the plasma velocity, $p$ is the plasma pressure, $\mu$ is the magnetic permeability, ${\bf B}$ is the magnetic field, ${\bf g}$ is gravitational acceleration, $\gamma$ is the ratio of specific heats which we take to be $5/3$, $R$ is the universal gas constant, $T$ is the temperature, and $\widetilde{\mu}$ is the mean molecular weight which takes a value of $0.6$ in the solar atmosphere.  

\subsection{{Basic Equilibrium}}

A very simple, uniform, vertical equilibrium magnetic field, ${\bf B}_0=\left(0,0,B_0\right)$, is chosen and the background pressure and density can be calculated from the momentum equation under equilibrium conditions of uniform temperature ($T_0$) and the ideal gas law.  
\begin{eqnarray*}
p_0\left(z\right)&=&p_0\left(0\right)e^{-z/H},\\
\rho_0\left(z\right)&=&\rho_0\left(0\right)e^{-z/H},
\end{eqnarray*}
where $H=RT_0/\left(\widetilde{\mu}g\right)$ is the coronal scale height ($\approx60$Mm).  Thus the plasma $\beta$ ($=2\mu p_0/B_0^2$) is dependent on $z$, ensuring that our waves will cross the $\beta\approx1$ layer.  

To study the nature of this propagation we use the linearised MHD equations.  Using subscript $0$ for equilibrium quantities and subscript $1$ for perturbed quantities we have the linearised equation of motion
\begin{equation}
\rho_0\frac{\partial{\bf v}_1}{\partial t}=-{\bf\nabla}p_1+\frac{1}{\mu}\left({\bf\nabla}\times{\bf B}_1\right)\times{\bf B}_0+\rho_1{\bf g},
\label{linmo.eqn}
\end{equation}
the linearised induction equation
\begin{equation}
\frac{\partial{\bf B}_1}{\partial t}={\bf\nabla}\times\left({\bf v}_1\times{\bf B}_0\right),
\label{linind.eqn}
\end{equation}
the linearised equation of mass continuity
\begin{equation}
\frac{\partial\rho_1}{\partial t}+{\bf\nabla}\cdot\left(\rho_0{\bf v}_1\right)=0,
\label{lincont.eqn}
\end{equation}
and the energy equation
\begin{equation}
\frac{\partial p_1}{\partial t}+\left({\bf v}_1\cdot{\bf\nabla}\right)p_0=\frac{\gamma p_0}{\rho_0}\left(\frac{\partial\rho_1}{\partial t}+\left({\bf v}_1\cdot{\bf\nabla}\right)\rho_0\right).
\label{linen.eqn}
\end{equation}
Henceforth it is assumed that we are working with the linearised equations and subscripts on the perturbed quantities are dropped.  

\subsection{{$x$-Dependence and Non-Dimensionalisation}}

To reduce our problem down to one dimension we assume that all variables are invariant in $y$, and choose the $x$-dependence to be given by trigonometric functions of $kx$
\begin{eqnarray*}
{\bf v}&=&\left(v_x\left(z,t\right)\sin{kx},0,v_z\left(z,t\right)\cos{kx}\right),\\
{\bf B}&=&\left(B_x\left(z,t\right)\sin{kx},0,B_z\left(z,t\right)\cos{kx}\right),\\
\rho&=&\rho\left(z,t\right)\cos{kx},\\
p&=&p\left(z,t\right)\cos{kx}.
\end{eqnarray*}

We also non-dimensionalise all variables by setting ${\bf v}=v_0\bar{\bf v}$, ${\bf B}=B_0\bar{B}$, $p=p_{00}\bar{p}$, $\rho=\rho_{00}\bar{\rho}$, $p_0=p_{00}\bar{p_0}$, $\rho_0=\rho_{00}\bar{\rho_0}$, $x=L\bar{x}$, $z=L\bar{z}$, $t=\tau\bar{t}$, and $k=\bar{k}/L$, where a bar denotes a dimensionless quantity and $v_0$, $B_0$, $p_{00}$, $\rho_{00}$, $L$, and $\tau$ are constants with the dimensions of the variable they are scaling.  We can then set $B_0/\sqrt{\mu\rho_0}=v_0$ and $v_0=L/\tau$, thus our speed is measured in units of $v_0$ which represents a constant background Alfv\'en speed.  We also set $\beta_0=2\mu p_{00}/B_0^2=2c_{S0}^2/\left(\gamma v_0^2\right)$ where $\beta_0$ is the plasma $\beta$ at a reference height of zero; note that $g=c_{S0}^2/\left(\gamma H\right)$, where $c_{S0}^2=\gamma p_{00}/\rho_{00}$.  Thus we generate the dimensionless versions of Equations~(\ref{linmo.eqn})\,--\,(\ref{linen.eqn}).  Under these scalings $\bar{t}=1$ (for example) refers to $t=\tau=L/v_0$; {\it i.e.~}the time taken for a wave to travel a distance $L$ at the reference background Alfv\'en speed.  The bar on quantities is now dropped and it is understood that we are working with dimensionless values.  

\subsection{{Linearised Equations}}

The linearised equations are
\begin{eqnarray}
\frac{1}{v_A^2\left(z\right)}\frac{\partial v_x}{\partial t}-\frac{\partial B_x}{\partial z}&=&\frac{kc_s^2}{\gamma}p+kB_z,\nonumber\\
\frac{1}{v_A^2\left(z\right)}\frac{\partial v_z}{\partial t}+\frac{c_s^2}{\gamma}\frac{\partial p}{\partial z}&=&-\frac{Lc_s^2}{H\gamma}\rho,\nonumber\\
\frac{\partial B_x}{\partial t}-\frac{\partial v_x}{\partial z}&=&0,\nonumber\\
\frac{\partial B_z}{\partial t}&=&-kv_x,\nonumber\\
v_A^2\left(z\right)\frac{\partial\rho}{\partial t}+\frac{\partial v_z}{\partial z}&=&\frac{L}{H}v_z-kv_x,\nonumber\\
v_A^2\left(z\right)\frac{\partial p}{\partial t}+\gamma\frac{\partial v_z}{\partial z}&=&\frac{L}{H}v_z-\gamma kv_x.\label{lineqns.eqn}
\end{eqnarray}
These may then be combined to form two second-order wave equations
\begin{eqnarray}
\left(v_A^2\left(z\right)\frac{\partial^2}{\partial z^2}-\left(c_s^2+v_A^2\left(z\right)\right)k^2-\frac{\partial^2}{\partial t^2}\right)v_x=kc_s^2\left(\frac{\partial}{\partial z}-\frac{L}{\gamma H}\right)v_z,\nonumber\\
\left(c_s^2\frac{\partial^2}{\partial z^2}-\frac{Lc_s^2}{H}\frac{\partial}{\partial z}-\frac{\partial^2}{\partial t^2}\right)v_z=-kc_s^2\left(\frac{\partial}{\partial z}-\frac{L}{H}\left(1-\frac{1}{\gamma}\right)\right)v_x,\label{wav.eqns}
\end{eqnarray}
which are in agreement with Ferraro and Plumpton (1958).  This form of the equations is much easier to analyse with the WKB approximation, as is done in Section~\ref{largek.sec}.  

\section{Numerical Simulations}
\label{Num Sims}

We numerically solve Equations~(\ref{lineqns.eqn}) using the MacCormack method.  This is a two-step, predictor-corrector, Lax-Wendroff method which is second-order accurate in time and space and, for linear harmonic waves, not strongly affected by numerical dispersion or diffusion.  As we wish to study the propagation of waves incident from above, we drive $v_z$ on the upper boundary, where we apply the conditions
\begin{eqnarray*}
v_x&=&0,\\
v_z&=&\sin{\omega t},\\
B_x&=&-\frac{1}{k}\frac{\partial B_z}{\partial z},\\
\frac{\partial B_z}{\partial t}&=&0,\\
\frac{\partial p}{\partial z}&=&0,\\
\frac{\partial\rho}{\partial z}&=&0.
\end{eqnarray*}
Imposing $v_z$ on the upper boundary means that we are predominantly driving a slow wave.  Since the slow wave also has a small component of $v_x$, the condition $v_x=0$ means that there is a small component of the fast mode generated; however, this mode is evanescent and does not propagate.  To ensure that our code is as accurate as possible, we use backward differencing for the predictor steps and forward differencing for the corrector steps; thus we are using the corrected values on the upper boundary.  The simulations are run for $-10\le z\le5$ and $0\le t\le35$; the end time is chosen so that we stop the simulation just before the wave front hits the lower boundary thus eliminating reflection effects.  In all simulations, we choose $\beta_0=0.2$ and $L$ equal to the coronal scale height ($H$) so that $z=1$ corresponds to one coronal scale height.  Having set these values, we are left with two free parameters: the driving frequency ($\omega$) and the wavenumber in the $x$-direction, $k$.  By altering these parameters we can make comparisons between our results and different analytical models.  

\subsection{{Wave Properties}}

 As we wish to investigate wave propagation across the $\beta\approx1$ layer we shall describe the properties of waves above and below this region.  An uncoupled slow magnetoacoustic wave ($k=0$ limit) propagating through a low-$\beta$ plasma will change its behaviour to that of a fast magnetoacoustic wave as it passes into high-$\beta$ plasma.  Similarly an uncoupled fast wave will change its behaviour to that of a slow wave as it travels from low to high-$\beta$ plasma.  Despite this change in behaviour, the wave mode is the same, no conversion has occurred.  Thus, when we discuss mode conversion, the slow wave driven on the upper boundary retains the properties of a slow wave as it propagates  down into the high-$\beta$ plasma.  We do not see any evidence of upward-propagating fast waves from the mode-conversion region.  The transmitted part of the incident slow wave will continue into the high-$\beta$ plasma as a fast wave.  

\begin{figure}[!p]
\centering
\includegraphics[scale=0.66]{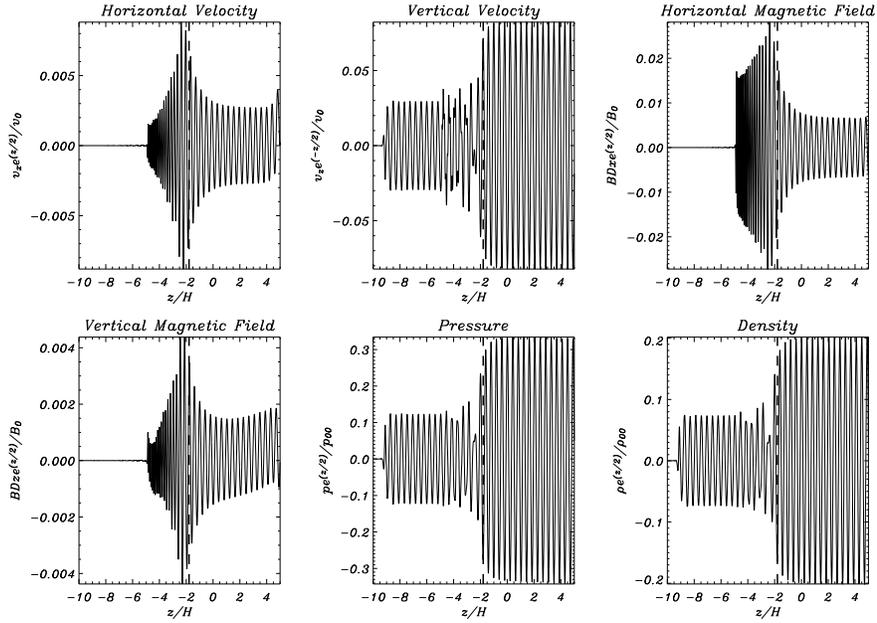}
\caption{Transformed variables at $t=35$, $k=\pi$, and $\omega=2\pi$.  The $c_s=v_A$ layer is denoted by a dashed line.}
\label{trans.fig}
\end{figure}

This is demonstrated by the numerical simulations seen in Figure~\ref{trans.fig}, which shows our variables under the transformations $v_x\rightarrow\widetilde{v}_xe^{-z/2}$, $v_z\rightarrow\widetilde{v}_ze^{z/2}$, $B_x\rightarrow\widetilde{B}_xe^{-z/2}$, $B_z\rightarrow\widetilde{B}_ze^{-z/2}$, $p\rightarrow\widetilde{p}e^{-z/2}$, and $\rho\rightarrow\widetilde{\rho}e^{-z/2}$.  These transformations remove the exponential behaviour, which is due to gravity, from the plots making other effects much clearer.  The wave driven in on the upper boundary is, as discussed, predominantly a slow wave and this behaviour is retained until the wave reaches the $c_s=v_A$ layer.  At this point, mode conversion occurs, and in the high-$\beta$ plasma (to the left of the dashed line) the converted part of the wave propagates as a slow wave and the transmitted part as a fast wave.  The plots of $\widetilde{v}_z$, $\widetilde{p}$ and $\widetilde{\rho}$ in Figure~\ref{trans.fig} clearly show the fast wave out in front at $z\approx-9.3$, the interference seen for $-4.9\le z\le-1.8$ is due to the interaction of the fast and slow waves.  The plots for $\widetilde{v}_x$, $\widetilde{B}_x$ and $\widetilde{B}_z$ illustrate that the interference stops at $z\approx-4.9$ because this is as far as the slow wave has propagated.  Gravity does not seem to affect the position of the mode conversion layer, which was also found to lie at $c_s=v_A$ in a weakly non-uniform medium.  

\begin{figure}[!p]
\centering
\includegraphics[scale=0.66]{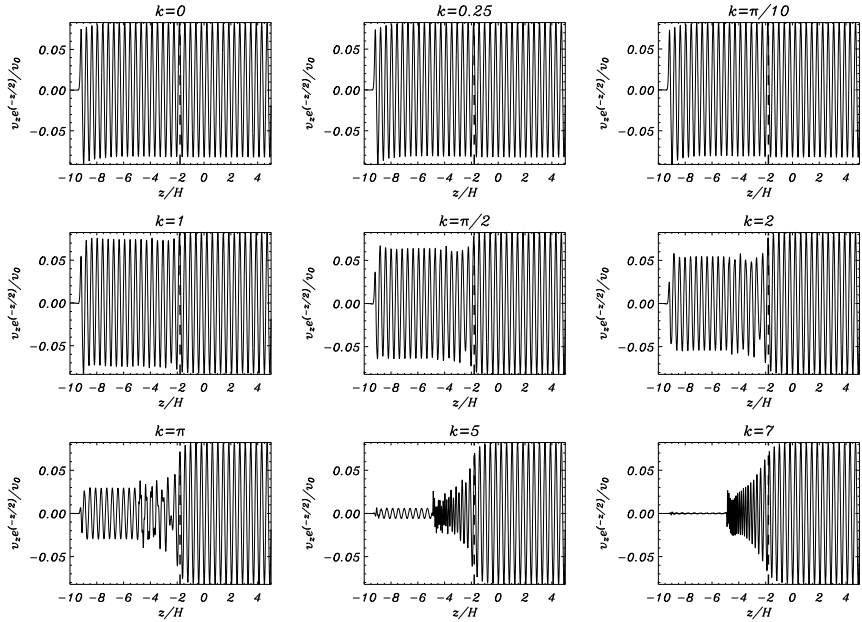}
\caption{Transformed vertical velocity ($\widetilde{v}_z$) for $t=35$ and $\omega=2\pi$.  The $c_s=v_A$ layer is denoted by a dashed line.}
\label{varyk2pi.fig}
\end{figure}

\subsection{{Varying the Wavenumber}}

We now consider the effect of varying the wavenumber, $k$, on the mode conversion.  Figure~\ref{varyk2pi.fig} shows the transformed vertical velocity, $\widetilde{v}_z$, at $t=35$ for a range of values for $k$.  The driving frequency, $\omega=2\pi$, corresponds to driving in a wave with a period of one Alfv\'en time, or $60$s.  

It is clear from the wave Equations~(\ref{wav.eqns}) that when $k=0$ the fast and slow magnetoacoustic waves are completely decoupled.  This can be observed in Figure~\ref{varyk2pi.fig} where no mode conversion is seen for $k=0$, and indeed is not even visible for very small $k$.  As $k$ increases further the amount of mode conversion also increases, until almost none of the incident wave can be seen beyond the $c_s=v_A$ layer for $k=7$.  In Section~\ref{smallk.sec} we quantify the proportion of mode conversion using an approximation valid for small $k$, and in Section~\ref{largek.sec} we investigate behaviour for large $k$.  

\subsection{{Small $k$}}
\label{smallk.sec}

Cairns and Lashmore-Davies (1983, 1986) and Cairns and Fuchs (1989) developed a method of solving mode-conversion problems, which derives differential equations describing the coupled mode amplitudes from the local dispersion relation.  These equations can then be solved analytically to find the transmission and conversion coefficients.  In relation to our simulations this method can be applied when $k$ is small, and $\omega$ is sufficiently large compared to $k$.  

Starting with Equations~(\ref{wav.eqns}) we take the time variation as $e^{i\omega t}$ so that $\partial/\partial t=i\omega$ and let $v_z=e^{z/2}V_z/c_s$.  Under the assumption that $k$ is small and $\omega$ large, the derivatives of $V_z$ and $v_x$ with respect to $z$ are larger than the functions themselves.  Thus, our equations may be reduced to
\begin{eqnarray*}
\left(\frac{d}{dz}+\frac{i\omega}{v_A\left(z\right)}\right)\left(\frac{d}{dz}-\frac{i\omega}{v_A\left(z\right)}\right)v_x&=&\frac{kc_s}{v_A\left(z\right)}\frac{dV_z}{dz},\\
\left(\frac{d}{dz}+\frac{i\omega}{c_s}\right)\left(\frac{d}{dz}-\frac{i\omega}{c_s}\right)V_z&=&-\frac{kc_s}{v_A\left(z\right)}\frac{dv_x}{dz}.
\end{eqnarray*}
The equation for $v_x$ is driven by the $V_z$ that is imposed at the upper boundary.  This inhomogeneous term has a wavenumber in $z$ given by $\omega/c_s$ and at $z_0$, where $v_z\left(z_0\right)=c_s$, there is a resonance and the amplitude of $v_x$ increases rapidly while the amplitude of $V_z$ is reduced.  This is mode conversion.  

Expanding $z=z_0+\xi$ around the mode conversion region ({\it i.e.~}for the downward propagating waves described by the brackets containing minus signs) and letting $d/dz=i\omega/c_s$ away from this region, our equations may be written
\begin{eqnarray}
\frac{dv_x}{d\xi}-i\left(\frac{\omega}{c_s}-\frac{\omega}{2c_s}\xi\right)v_x&=&\frac{kc_s}{2v_A\left(z\right)}V_z,\nonumber\\
\frac{dV_z}{d\xi}-\frac{i\omega}{cs}V_z&=&-\frac{kc_s}{2v_A\left(z\right)}v_x.\label{Cairns.eqn}
\end{eqnarray}
These satisfy the condition for energy conservation
\begin{displaymath}
\frac{d}{d\xi}\left(\left|v_x\right|^2+\left|V_z\right|^2\right)=0.
\end{displaymath}
Eliminating $v_x$ from Equations~(\ref{Cairns.eqn}) and making the transformations given by Cairns and Lashmore-Davies
\begin{eqnarray*}
V_z\left(\xi\right)&=&\exp\left(\frac{i\omega}{c_s}\xi-\frac{i\omega}{8c_s}\xi^2\right)\psi\left(\xi\right),\\
\zeta&=&\left(\frac{\omega}{2c_s}\right)^{\frac{1}{2}}\xi e^{3i\pi/4},
\end{eqnarray*}
we obtain
\begin{equation}
\frac{d^2\psi}{d\zeta^2}-\left(\frac{\zeta^2}{4}-\frac{ik^2c_s}{2\omega}-\frac{1}{2}\right)\psi=0.
\label{parcyl.eqn}
\end{equation}
The solution of this equation is given by the parabolic cylinder function $U\left(a,\zeta\right)$ where
\begin{displaymath}
a=-\frac{ik^2c_s}{2\omega}-\frac{1}{2}.
\end{displaymath}
The asymptotic behaviour of these functions is described in detail in Abramowitz and Stegun (1964).  

\begin{figure}
\centering
\includegraphics[scale=0.71]{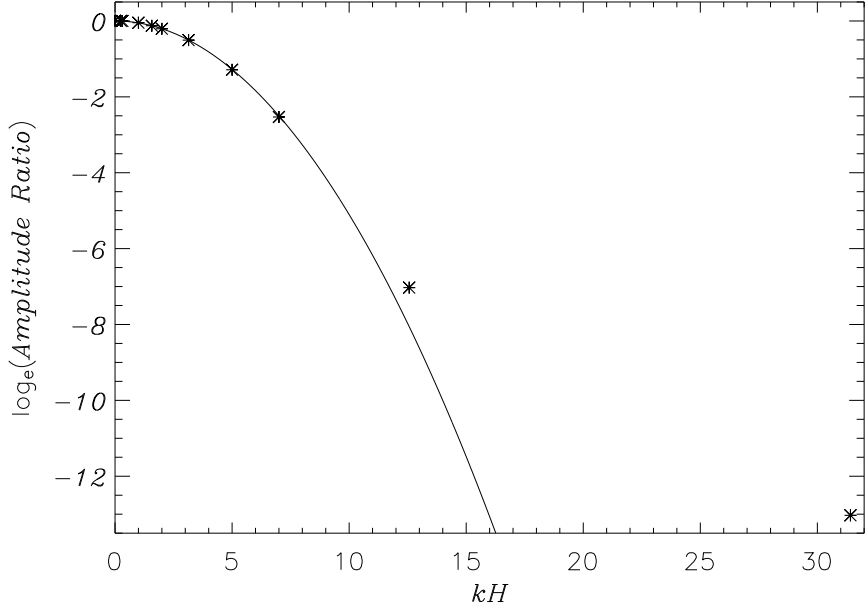}
\caption{The logarithm of the ratio of slow wave amplitudes in high and low $\beta$.  The solid line gives the values predicted using Equation~(\ref{amp.eqn}) and the stars are values calculated from numerical simulations run for $\omega=4\pi$.}
\label{logamp.fig}
\end{figure}

Equations~(\ref{Cairns.eqn}) may also be compared directly to the expressions used in Cairns and Lashmore-Davies (1983) to find the energy transmission and so the ratio of wave amplitudes in high and low-$\beta$ is given by
\begin{equation}
A=\exp\left(-\frac{\pi k^2c_s}{2\omega}\right).
\label{amp.eqn}
\end{equation}
Thus, if we know the amplitude of the incident slow wave, we may multiply by $A$ to predict the amplitude of the transmitted fast wave.

A comparison between this analytical approximation and our numerical data is shown in Figure~\ref{logamp.fig}.  The logarithm has been taken to highlight that the data is in excellent agreement for small $k$; no visible difference could be seen between the analytical prediction~(\ref{amp.eqn}) and the original numerical data.  

Cally (2006) and Schunker and Cally (2006) apply ray tracing theory to the mode conversion problem.  They also find these transmission coefficients.  

\subsection{{Large $k$}}
\label{largek.sec}

For large $k$ we may compare our numerical simulations with an analytical approximation derived by Roberts (2006).  In the limit of large $k$ the first of our wave equations~(\ref{wav.eqns}) reduces to
\begin{equation}
\left(c_s^2+v_A^2\left(z\right)\right)kv_x+c_s^2\left(\frac{\partial}{\partial z}-\frac{1}{\gamma}\right)v_z=0.
\label{vxapprox.eqn}
\end{equation}
We may use this to eliminate $v_x$ from the first wave equation giving
\begin{displaymath}
\frac{\partial^2v_z}{\partial t^2}-c_T^2\left(z\right)\frac{\partial^2v_z}{\partial z^2}+\frac{c_T^4\left(z\right)}{c_s^2}\frac{\partial v_z}{\partial z}+\frac{c_s^2c_T^2\left(z\right)}{\gamma v_A^2\left(z\right)}\left(\frac{c_T^2\left(z\right)}{c_s^2}-\left(\frac{1}{\gamma}-1\right)\right)v_z=0,
\end{displaymath}
where $1/c_T^2\left(z\right)=1/c_s^2+1/v_A^2\left(z\right)$ is the tube speed.  
If we introduce
\begin{equation}
Q\left(z,t\right)=\left(\frac{\rho_0\left(z\right)c_T^2\left(z\right)}{\rho_0\left(0\right)c_T^2\left(0\right)}\right)^{\frac{1}{2}}v_z\left(z,t\right),
\label{kleintrans.eqn}
\end{equation}
then our equation can be written in the form of the Klein-Gordon equation
\begin{equation}
\frac{\partial^2Q}{\partial t^2}-c_T^2\left(z\right)\frac{\partial^2Q}{\partial z^2}+\Omega^2\left(z\right)Q=0,
\label{klein.eqn}
\end{equation}
where
\begin{eqnarray*}
\Omega^2\left(z\right)&=&c_T^2\left(z\right)\left\{\frac{1}{4}\frac{c_T^4\left(z\right)}{c_s^4}-\frac{c_T^4\left(z\right)}{2c_s^2v_A^2\left(z\right)}+\frac{c_s^2}{\gamma v_A^2\left(z\right)}\left(\frac{c_T^2\left(z\right)}{c_s^2}-\left(\frac{1}{\gamma}-1\right)\right)\right\},
\end{eqnarray*}
is a cut-off frequency.  

Using the Klein-Gordon equation~(\ref{klein.eqn}) with the term involving the cut-off frequency neglected, and the formal WKB method, as described in Bender and Orszag (1978), we find a solution of $Q\left(z,t\right)$, valid for $\omega$ large, to leading order
\begin{eqnarray*}
Q\left(z,t\right)&\sim&Dc_T^{\frac{1}{2}}\left(z\right)\sin\left\{\omega\left[t+\int_z^{z_{m}}\frac{1}{c_T\left(z\right)}dz\right]\right\}\\&=&Dc_T^{\frac{1}{2}}\left(z\right)\sin \left\{\omega\left[t+\frac{1}{c_s}\ln\left|\frac{\left(c_T\left(z\right)+c_s\right)\left(c_T\left(z_{m}\right)-c_s\right)}{\left(c_T\left(z\right)-c_s\right)\left(c_T\left(z_{m}\right)+c_s\right)}\right|\right.\right.\\&&\left.\left.-2\left(\frac{1}{c_T\left(z\right)}-\frac{1}{c_T\left(z_{m}\right)}\right)\right]\right\},
\end{eqnarray*}
where $D$ is a constant and $z_{m}$ is the maximum value of $z$.  We can then find an expression for $v_z\left(z,t\right)$ using Equation~(\ref{kleintrans.eqn})
\begin{eqnarray}
v_z\left(z,t\right)&=&\frac{v_A\left(z\right)c_T^{1/2}\left(z_{m}\right)}{v_A\left(z_{m}\right)c_T^{1/2}\left(z\right)}\sin\left\{\omega\left[t+\frac{1}{c_s}\ln\left|\frac{\left(c_T\left(z\right)+c_s\right)\left(c_T\left(z_{m}\right)-c_s\right)}{\left(c_T\left(z\right)-c_s\right)\left(c_T\left(z_{m}\right)+c_s\right)}\right|\right.\right.\nonumber\\&&\left.\left.-2\left(\frac{1}{c_T\left(z\right)}-\frac{1}{c_T\left(z_{m}\right)}\right)\right]\right\}.
\label{analvz.eqn}
\end{eqnarray}

\begin{figure}[!b]
\centering
\includegraphics[scale=0.66]{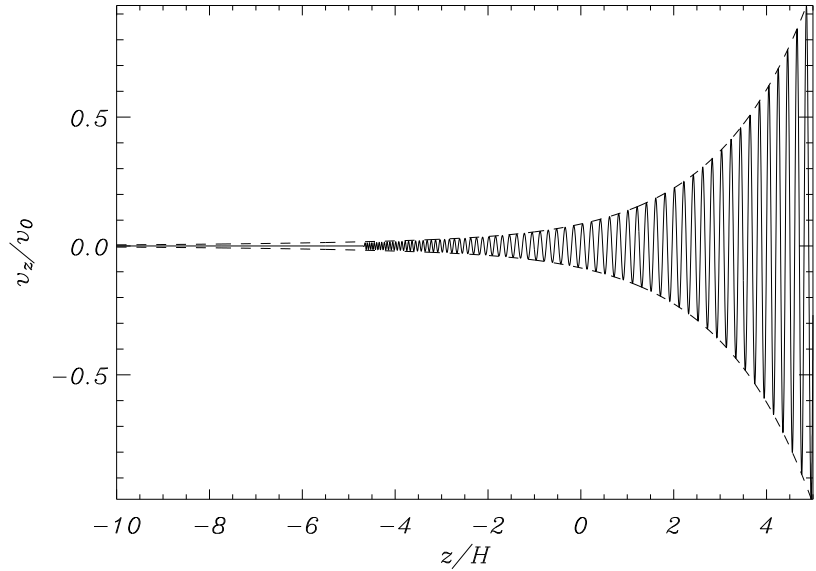}
\caption{Vertical velocity for $\omega=4\pi$ and $k=40\pi$ at $t=35$ (solid) and wave amplitude predicted from Equation~(\ref{analvz.eqn}) (dashed).}
\label{ampvz.fig}
\end{figure}

\begin{figure}[p]
\centering
\includegraphics[scale=0.66]{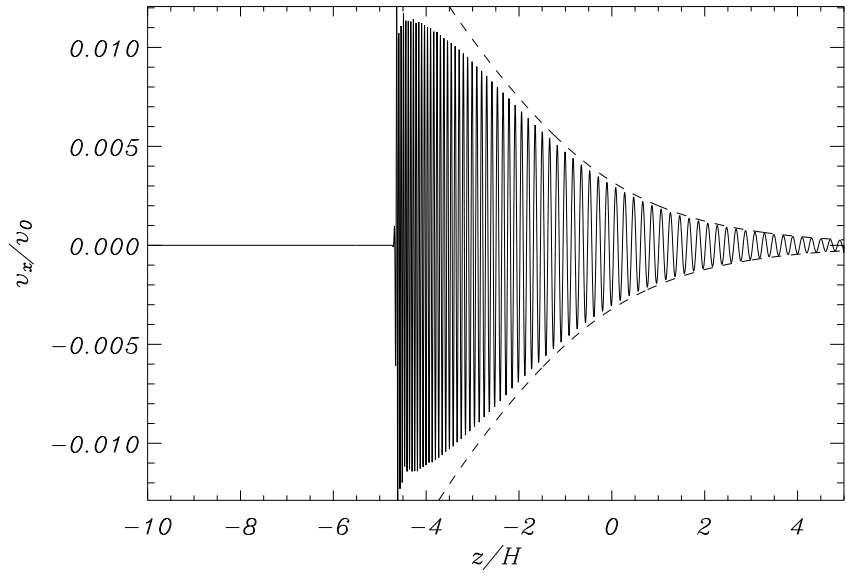}
\caption{Horizontal velocity for $\omega=4\pi$ and $k=40\pi$ at $t=35$ (solid) and wave amplitude predicted using Equations~(\ref{vxapprox.eqn}) and~(\ref{analvz.eqn}) (dashed).}
\label{ampvx.fig}
\end{figure}

\begin{figure}[p]
\centering
\includegraphics[scale=0.66]{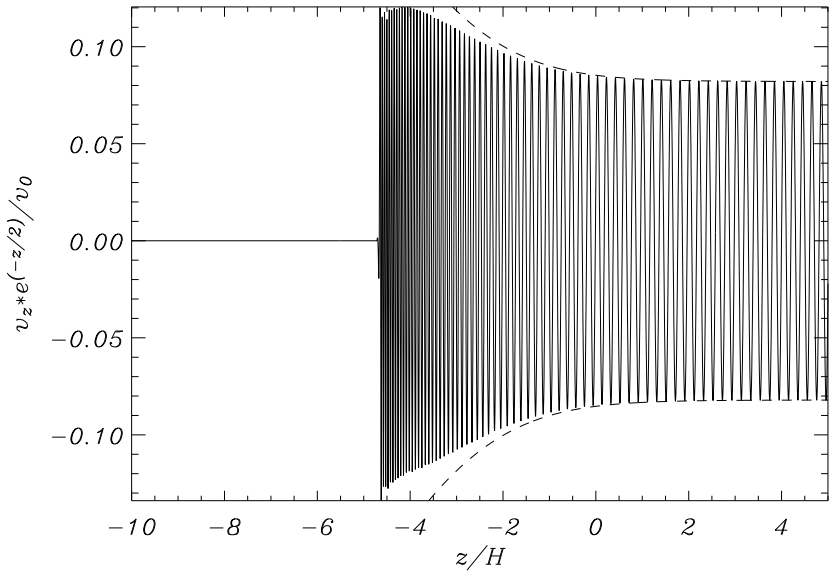}
\caption{Transformed vertical velocity, $\widetilde{v}_z$, for $\omega=4\pi$ and $k=40\pi$ at $t=35$ (solid) and wave amplitude predicted using Equation~(\ref{analvz.eqn}) (dashed).}
\label{transampvz.fig}
\end{figure}

This expression agrees extremely well with the data obtained from the numerical simulations, as can be seen in Figure~\ref{ampvz.fig}.  An analytical approximation to $v_x$ can then be calculated using Equation~(\ref{vxapprox.eqn}); a comparison of this result with the numerical data is shown in Figure~\ref{ampvx.fig}.  The agreement between the numerical simulations and the analytical approximation is clearly not as good for $v_x$ (Figure~\ref{ampvx.fig}) as it is for $v_z$ (Figure~\ref{ampvz.fig}).  If we remove the exponential behaviour from the vertical velocity, as is shown in Figure~\ref{transampvz.fig}, we can see that for $z<0$ the agreement between the analytical and numerical results worsens.  This is the same area for which we begin to see a deviation between the two solutions for the horizontal velocity.  In order to improve agreement a higher-order WKB solution should be found.  

\section{Conclusions}
\label{conc.sec}

In this paper we have considered the downward propagation of linear waves in an isothermal atmosphere permeated by a vertical background magnetic field.  More specifically, we concentrated on the region where the wave passes from low-$\beta$ to high-$\beta$ plasma.  As expected our simulations show mode conversion occurring as the wave passes through this region at the point where the sound and Alfv\'en speeds are equal (Figure~\ref{trans.fig}), similar to a weakly non-uniform medium.  One of the free parameters in our simulations is the horizontal wavenumber ($k$).  Figure~\ref{varyk2pi.fig} shows that the value of $k$ has a strong effect on the amount of mode conversion that occurs.  For $k=0$, the two wave modes are completely decoupled and no mode conversion occurs, and as $k$ increases so does the degree of mode conversion.  As the wavenumber was shown to be an important factor, we then looked at analytical approximations in the limit of both small and large $k$.  

For small $k$ we applied the method developed by Cairns and Lashmore-Davies (1983) for analysing mode conversion.  This allowed us to find a solution for the vertical velocity ($v_z$) which is valid in the mode-conversion region.  This solution is related to the parabolic cylinder function, for which asymptotic solutions are readily available \cite{Abram1964}.  We also found the amplitude of the transmitted wave, predicted from the amplitude of the incident wave, is in excellent agreement with our numerical data (Figure~\ref{logamp.fig}).  This equation backs up our observation that the mode conversion increases with $k$ and reveals that $\omega$ also has an effect on mode conversion, namely mode conversion decreases as $\omega$ increases.  However, the effect due to the wavenumber $k$ is the stronger of the two.  

For large $k$ we followed an analysis carried out by Roberts (2006).  From this we found a WKB solution for $v_z$ valid for large $\omega$ (Equation~(\ref{analvz.eqn})).  This in turn was used to find an analytical solution for $v_x$ using Equation~(\ref{vxapprox.eqn}).  Figure~\ref{ampvz.fig} shows very good agreement between the analytical and numerical results for the vertical velocity, however once the exponential behaviour is removed (Figure~\ref{transampvz.fig}) we see that the analytical solution begins to deviate from the numerical results for $z<0$.  This is also true of the horizontal velocity ($v_x$) shown in Figure~\ref{ampvx.fig}.  Equation~(\ref{vxapprox.eqn}) linking $v_x$ to $v_z$ itself is a very good approximation provided $k$ is large.  

A thorough investigation of this very simple one-dimensional model has yielded some very interesting results, giving us some insight into the mode conversion problem.  We hope to extend this work to include a non-isothermal atmosphere, making the model more physically realistic.  

\acknowledgements

Dee McDougall acknowledges financial assistance from the Carnegie Trust for the Universities of Scotland.

\end{article} 
\end{document}